# The Iberian Exception:
# An overview of its effects over its first 100 days

D. Robinson[1], A. Arcos-Vargas[2], M. Tennican[3], F. Núñez[4]

September 25, 2023

## Abstract

This paper offers an independent assessment of certain key economic effects of the Iberian Exception (IE), the common name for legal measures affecting the Iberian power market that were introduced in June 2022 by the Spanish and Portuguese governments. Their stated aim was to reduce the major component of electricity prices for most Iberian consumers, a component which was indexed to Iberian wholesale power market spot prices—power market prices that were rising alarmingly due to extremely tight international markets for natural gas. According to the Iberian governments, this objective was to be attained by terms of the IE that subsidize a reduction in wholesale power market prices, with that subsidy financed in part by a new element added to the bills of consumers benefiting from that wholesale price reduction. Another Spanish governmental aim was to reduce the government's published measure of inflation, which was linked to a regulated retail price indexed to Iberian wholesale spot power market prices.

The Spanish Government estimates that, during its first 100 days, the IE provided substantial benefits for consumers affected by the IE, which included over 10 million small consumers as well as many large ones, but the authors of this study question that estimate. The authors of this paper argue that the estimated effect of the IE on retail prices depends critically on the assumptions about what would have occurred in the absence of the IE, i.e., in a counterfactual scenario. Although counterfactuals are always difficult to construct, the government´s counterfactual ignores demand elasticity, and this inflates their estimate of immediate consumer benefits. Using hourly data on the wholesale electricity market for the first 100 days of the IE, this paper's analysis of alternative counterfactuals that reflect the effects of demand elasticity shows substantially lower benefits of the IE for consumers than the Spanish government estimates. Indeed, this paper's analysis suggests that affected consumers would have paid somewhat less for electricity in the first 100 days of the IE had it not been introduced.

The authors identify several other potential short- and long-term effects of the IE that deserve further study. These include increased margins for fossil fired generators, reduced margins for some decarbonized inframarginal plant, heightened investor perceptions of regulatory risk, weakened incentives for efficient consumption, and higher carbon emissions and gas prices.

[1] Oxford Institute for Energy Studies
[2] Oxford Institute for Energy Studies, Universidad de Sevilla
[3] Tennican & Associates
[4] Universidad de Sevilla

# Introduction and Summary

In June 2022, the Spanish and Portuguese governments introduced parallel laws, commonly called the Iberian Exception (IE), to reduce wholesale electricity prices by decoupling them from wholesale natural gas prices, which had been rising sharply since the first quarter of 2021. The IE was also a Spanish governmental response to its own circumstances, notably the fact that the energy component of regulated retail power prices of many Spanish consumers was directly indexed to Iberian wholesale spot prices and the further fact that these retail prices were a key input to Spain's official inflation index. At the time, the Iberian market intervention was contrary to general European Union (EU) policy but was approved as an exception by the European Commission after the Spanish and Portuguese governments agreed that its intended reduction in Iberian wholesale market prices would apply also to export sales of power, notably to France.

The IE led to a reduction in wholesale Iberian power market prices primarily by offering an out-of-market subsidy to fossil-fired generation, notably gas-fired generation.[5] These generators reduced their offers to reflect the subsidy they would receive, resulting in lower power-market clearing prices than would otherwise have occurred. This subsidy is called the generation contribution (GC). It was funded primarily by requiring the affected Iberian consumers with electricity prices indexed directly to wholesale power prices, to make a demand contribution (DC); the affected consumers included over 10 million small consumers in Spain. Customers with power purchase contracts on fixed prices were initially not required to pay the DC. However, with the annual renewal of fixed price contracts, many in Spain were deemed also to be benefiting from lower wholesale power market prices and therefore had to pay the DC, often in addition to the fixed price they had already agreed on multi-year contracts.

The Spanish Government (including the Ministry responsible for energy[6]) maintain that the IE significantly reduced the cost of electricity for affected consumers during the first few months[7]. The authors have found no official public document outlining the methodological basis for the Ministry's calculation. However, their understanding is that the Ministry´s calculations are based on a counterfactual that assumes no change in demand when wholesale prices fall. Various authors have used that methodology to estimate the savings for affected consumers. For instance, Sancha[8] estimates savings of about 18% for the period from 15 June to the end of September 2022, which is slightly longer than the first 100 days. This paper refers to that methodology as the Spanish Government's methodology.

---

[5] The IE also contained other terms affecting hydro, nuclear, wind and solar facilities. The subsidy to fossil-fired plants included gas and coal fired generation during the first 100 days. It was extended to cogeneration, but this occurred after the 100-day period.

[6] The Ministry for the Ecological Transition and the Demographic Challenge.

[7] See https://www.lamoncloa.gob.es/presidente/actividades/Paginas/2022/060922-sanchez-senado.aspx; and https://www.europapress.es/economia/energia-00341/noticia-ribera-estima-excepcion-iberica-representa-ahorro-17-euros-mes-recibo-luz-20220928101422.html; https://www.libremercado.com/2022-08-17/teresa-ribera-atribuye-la-reduccion-del-consumo-energetico-a-sus-medidas-y-no-al-fin-de-la-ola-de-calor-6924217/

[8] José Luis Sancha, *"Balance económico de cuatro meses de aplicaciones del tope al precio de gas, Cuadernos de Energía"*, *Numero* 70, 7 *noviembre* 2022.

Drawing on detailed hourly data published by the Iberian market operator (OMIE) for the first 100 days of the IE implementation, this paper questions the Spanish Government´s methodology for estimating consumer benefits. It argues that this methodology ignores the effect of the IE on electricity demand, in particular the increase in power flows over the interconnector between Spain and France. The lower Iberian wholesale power market prices resulting from the IE significantly increased exports from Spain to France and may also have increased Iberian consumer demand. Higher exports and domestic demand increased Iberian power generation and Iberian wholesale power market costs and prices by amounts that the Ministry's methodology apparently ignores. The analysis summarized in this paper demonstrates that the net benefits for Spanish consumers were less than the Government suggests. Under one scenario presented in this paper, the energy component of electricity prices for these consumers during the period studied may indeed have been higher under the IE than had there been no such governmental intervention. Given the importance of the energy component during the period studied a focus on the impact of the IE is justified. Furthermore, during its first 100 days, the IE contributed to a significant increase in gas-fired generation and consequent carbon emissions in Iberia. Finally, the analysis suggests that increasing demand led to increased margins for fossil-fired plant.

The authors stipulate that their analysis, like the Ministry's claims, necessarily is based on assumptions about what would have happened in the absence of the IE, that is, in a counterfactual scenario. Although they are convinced that their counterfactual assumptions and analytical methodology are more realistic and sounder than those of the Government, they recognize that their assumptions and analysis are open to challenge and refinement. Nevertheless, their analysis suggests that the Ministry´s estimates of the benefits of the IE for Spanish consumers in the first 100 days are misleading.

Although this paper focuses on short-run and quantifiable consequences of the IE, the introduction of the IE introduced other distortions with likely longer-term economic consequences. The first is that the IE seems likely to have reinforced Spain's reputation for regulatory risk. Second, the IE redirected revenues and profits toward fossil-fired power plants and away from decarbonized generation, a direction seemingly inconsistent with the Iberian Governments' stated energy transition objectives. Furthermore, to the extent that consumers believed that the IE significantly reduced their net power costs (which was certainly the case for French imports), the Ministry will have discouraged consumer investments in the efficient use of energy (e.g., installing heat pumps or power storage capacity), indirectly contributing to higher demand and prices for electricity and gas than would otherwise have occurred.

Since the IE has been extended to the end of 2023, and may be extended again, further work is needed to buttress and extend the analysis of the IE's consequences and to develop other and better approaches to manage the consequences of extreme natural gas prices for Spain and Portugal. Although gas prices currently have been lower than during the period studied, the authors believe it would be advisable to be prepared for possible future accidents, strikes or strategic withholdings that could again curtail gas supply and inflate gas prices.

This paper has seven sections, in addition to this introduction.

- The first is an introduction to the Iberian electricity market (MIBEL).
- The second introduces the Iberian Exception (IE) in more detail.
- The third introduces a conceptual model of the effects of the IE.
- The fourth analyzes the quantitative impact on prices for affected Spanish consumers for the first 100 days of the IE, using the Government´s counterfactual.
- The fifth section analyzes the impact on consumers over the same period under two alternative counterfactual scenarios that illustrate the importance of demand elasticity.
- The sixth section identifies several areas for further research.
- The seventh contains concluding comments.

## 1. The Iberian Electricity Market (MIBEL) Supply, Demand and Prices

MIBEL is the integration of Spanish and Portuguese wholesale electricity markets. OMIE (Iberian Energy Market Operator - Spanish Pole) manages the MIBEL spot market, which includes a daily market and six intraday markets. OMIP (Iberian Energy Market Operator - Portuguese Pole) manages the MIBEL derivatives market. Prices in the Spanish and Portuguese markets are usually the same. This paper focuses on the OMIE-managed daily market and will usually refer to the Iberian market, except in the case of trade between Spain and France. The interconnectors between these countries are managed by the national system operators and congestion rents shared 50/50 between them[9].

As in most countries with liberalized electricity systems, the Iberian wholesale electric power market is based on marginalist principles. Producers and consumers (or retail suppliers acting in their stead) make their energy supply and demand offers for each of the 24 hours of the following day[10]. When arrayed by price, these offers constitute supply and demand curves. The supply curve is upward sloping, i.e., more power will be offered and supplied by generators as market prices increase, and the demand curve is downward sloping, i.e., less power will be bid for and consumed as market prices increase. In conformity with basic economic logic, market equilibrium prices and volumes are set where the supply and demand curves cross. The market operator then dispatches generation units in ascending order of cost until the total of generation offers reaches the market equilibrium requirement.

Generators' offer prices in any hour are generally determined by their short-run marginal costs, which include variable operation and maintenance costs, but the marginal costs of fossil-fuel plants are dominantly the costs of the fuels consumed. In general, generators' bids are associated with their technologies, although this is not always the case since different cost structures and commitments may lead them to bid different values. In Iberia, gas-fired generation plants and thus their gas supply costs largely determine wholesale power market prices during most hours of most days. In the first hundred days of the IE, gas-fired combined cycle plants fixed the price 44% of the time, while manageable (pumped storage and conventional reservoir) hydro did so 41% of the time. (See Table 1 below.)

---

[9] As part of the IE, Spain agreed to share its 50% share of the French-Spanish congestion rents with all the consumers in Spain, Portugal and Morocco affected by the IE.

[10] In fact, there are seven markets with different time frames, but there is no need here to take separate account of them in this overview of the effects of the IE.

Even when the marginal technologies are hydro or other renewables, their offers are often determined by the price of gas-fired plants. In the case of manageable hydro, bidding in line with gas generation costs is usually due to the opportunity cost of the hydro resource, namely the alternative of using stored water later when gas generation is needed to meet demand and market-clearing prices, reflecting gas costs, are higher. For intermittent resources, like wind, solar and run-of-river hydro, generators' offer prices also sometimes reflect the cost of buying electricity in short-term imbalance markets to meet expected deviations from day-ahead commitments; these short-term market prices for imbalances also usually reflect the cost of natural gas fired plants.

| **Generation at the margin** | **Hours** | **%** | |
|---|---|---|---|
| Combined cycle gas turbine | 1,056 | 44% | |
| Pumped storage hydro | 277 | 12% | 41% |
| Conventional reservoir hydro | 701 | 29% | |
| Renewables, cogeneration & waste | 341 | 14% | |
| Conventional thermal (coal) | 25 | 1% | |
| | **2,400** | **100%** | |

**Table 1:** Generation at the margin in MIBEL over the first 100 days of the IE. Source: OMIE.

Generation offers are accepted and generation units are dispatched by the power system operator in merit order, from the lowest to the most expensive offer prices, but the owners of generation units do not necessarily receive the spot market clearing price. Indeed, most generation is covered by contracts (with retailers, final consumers, or financial markets), the details of which are beyond the scope of this paper. References here to generation revenues and costs ignore the distribution of economic consequences across ownership and contractual interests. They simply refer to the aggregate benefits and costs of each major type of generation in the Iberian power system using the wholesale spot market price and the generation contribution, if applicable, as a metric.

In the case of consumer demand and consumer offer prices, a distinction can be made between two kinds of demand. On the one hand, the demand of retailers supplying residential consumers can be viewed as vertical sections of the overall demand curve because their customers' demand is assumed to be price inelastic, at least in the very short run, i.e., essentially unaffected by power prices. Accordingly, retail suppliers offer prices that are set at a level that aims to ensure that their customers' demand is satisfied. In practice, this distinction may underestimate the elasticity of demand by smaller consumers, especially in the mid to long term. On the other hand, the demand of large customers (industrial, commercial, etc.) can be viewed as a price-elastic, downward sloping section of the overall demand curve, reflecting a decrease in the economic attractiveness of electric power use in their productive activities as power prices increase.

What is of critical importance is that export demand (over interconnectors) is, like the demand of some large Iberian consumers, price elastic and finely attuned, on an hourly or even shorter time scale, to the level of Iberian wholesale prices compared to foreign (most importantly, French) wholesale prices. When Spanish wholesale market prices are below French wholesale prices, Spain will export until prices are equal in both markets or until the interconnector is full, at which point the two systems share the congestion rents[11] 50/50. Conversely, generators in France can serve as a source of power supply to Iberia when French wholesale prices are lower than Spanish prices.

Using OMIE data, and for illustrative purposes, Figure 1 depicts actual supply and demand offers in the Iberian market relating to one hour during one day in the first 100 days of the IE.

**Figure 1.** Illustrative supply and demand curves for Iberian wholesale market

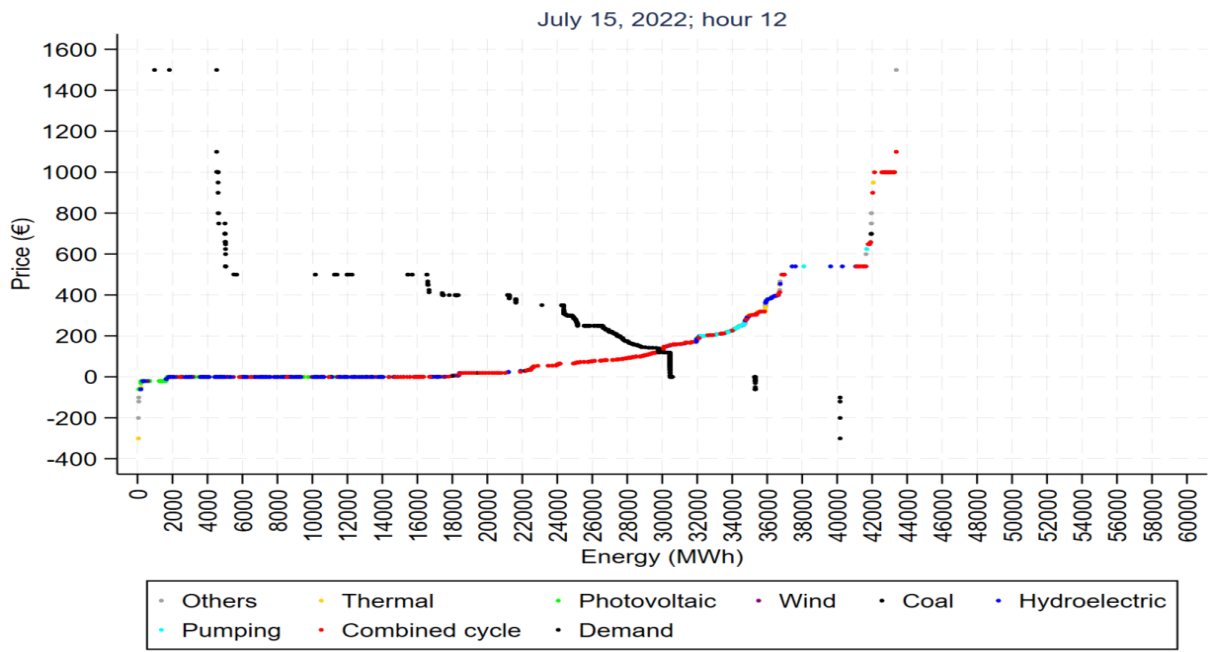

**Source:** Own elaboration based on OMIE data.

## 2. The Iberian Exception (IE)

The IE was designed to decouple the wholesale electricity price in the Iberian market from the wholesale market price of natural gas. The IE was introduced by Spanish Royal Decree-Law 10/2022 (and parallel Portuguese legislation) after a long negotiation with the European Commission (EC). It was approved for one-year but has since been extended by the EC until the end of 2023.

The stated Spanish and Portuguese governmental objective behind the IE was to reduce the wholesale price of electricity and consumers' power costs at a time when natural gas prices were very high and rising. Whereas in 2019 wholesale power costs were approximately €16 billion (37% of the total cost of electricity), in 2022 they were over €37 billion (61% of total cost), as can be seen in Figure 2. Since the energy component of Spain's regulated retail electricity tariff[12] for over 10 million consumers was indexed to the wholesale spot market, rising wholesale power prices were causing considerable social and political alarm in 2021 and 2022. Over time, the government reduced taxes and other charges to cushion the impact of high electricity costs on consumers, but these moves offset the effect of rising gas and wholesale power market prices only to a limited degree.

---

[11] Congestion rent (per kWh) is equal to the difference between the wholesale prices in the two markets when the network is congested, i.e., is at capacity.

**Figure 2.** Cost taxonomy of the Spanish electricity system 2018-2022

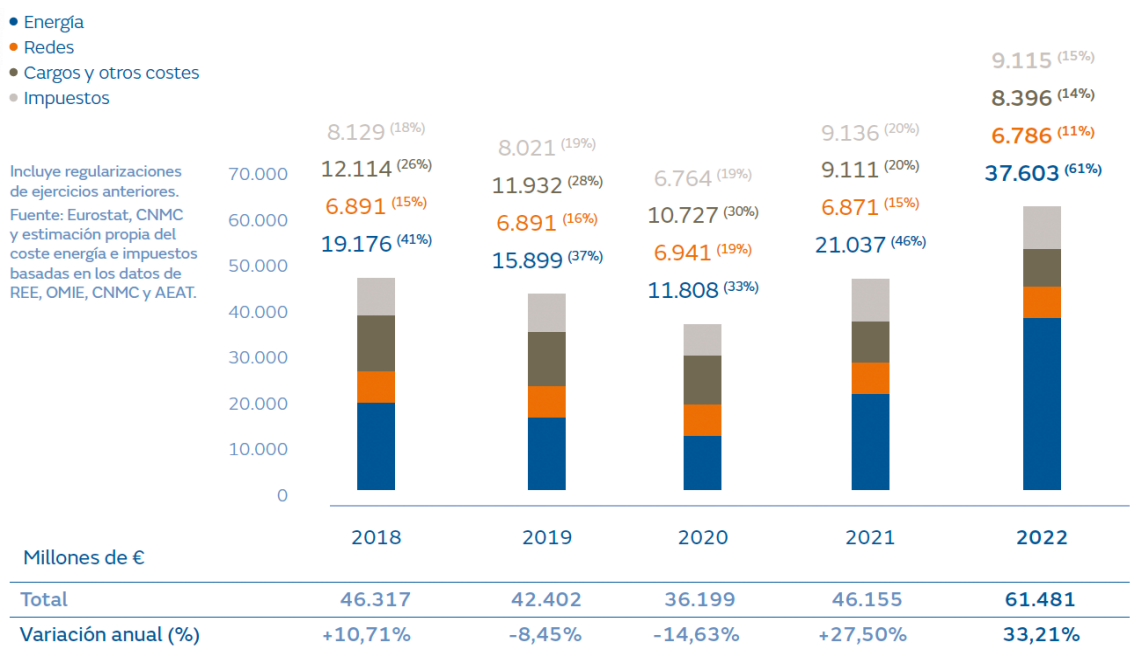

**Source:** Fundación Naturgy, 2023. El sector eléctrico español en números, Informe 2022, page 36.

It should be noted that the regulated retail electricity tariff was a key input to reported Spanish rates of inflation. Limiting wholesale power prices with the IE also addressed the Spanish government's concern about the country's high reported rates of inflation in 2021 and early 2022.

Beyond limiting the immediate impact of wholesale electricity prices on consumer electricity prices and published figures on inflation, the IE was also consistent with the Spanish Government´s view that the current European wholesale market design needed to be reformed. The view was that the cost and offer price of the marginal resource needed to meet demand (normally natural gas-fired generation) should not determine the market price for all the lower-cost, inframarginal hydro, renewables, and other generators.

---

[12] The PVPC (*Precio Voluntario para el Pequeño Consumidor*) is the regulated retail price for small consumers, of which there were more than 10 million when the IE was introduced.

The IE's reduction in wholesale power market prices was achieved by paying a subsidy to fossil-generators; the subsidy is referred to as a Generator Contribution (GC). The fossil generators, referred to here as the "privileged" or "affected" generators, adjusted their offer prices to reflect this subsidy, which is based on the difference between a new governmental wholesale gas reference price and the daily price of natural gas in the Iberian gas market (MIBGAS). The GC subsidy is intended to motivate generators to submit offer prices to the market operator that would be lower than prior to the IE by the amount of the GC, which would in turn lead to market-clearing prices also lower by the amount of the GC than would prevail in the absence of the IE. The lower market clearing prices on power generation received by affected generators under the IE in comparison to prices without the IE are, at least in theory, just offset by the generator contribution.

The gas reference price starts with a value of 40 €/MWh during the first six months of application, increasing subsequently by 5 €/MWh monthly, until it reaches 70 €/MWh at the end of the period[13]. However, gas prices are set by international gas market forces and generators are free under the IE to bid into the electricity market as they see fit. In practice, their bids were often higher than implied by the government´s estimate of gas prices because generators purchased gas in international markets, notably the TTF market, where gas prices were often higher than in the MIBGAS. As explained later in this paper, the higher cost of gas for some generation, along with higher demand for electricity, resulted in higher margins for some gas-fired plants and lower benefits for consumers under the IE than the government´s methodology suggests.

It is important to note that the GC is financed in large part by a new adjustment cost or demand contribution (DC) imposed by the IE on a large class of Iberian consumers[14]. These affected consumers are ones with power prices linked to wholesale power market prices and deemed by the Iberian governments to benefit from the reduction in the wholesale prices brought about by the IE. This adjustment cost or demand contribution (DC) is an important offset to any wholesale power market price reduction brought about by the IE, as is discussed in later sections of this paper. Under the IE, Spain's share of the congestion rents[15] obtained by the Spanish Transmission System Operator related to the cross-border electricity trade between France and Spain also were allocated to the financing of the GC, but as discussed later in this paper, these rents were relatively small compared to the demand contribution.

The amount of the DC compensation per kWh paid by affected Spanish and Portuguese consumers can vary depending on several factors. The DC will generally be greater:

---

[13] Following the EU extension of the IE until the end of 2023, the reference price will rise more slowly than initially planned.

[14] There are two parts to the adjustment cost, one of which refers to the higher costs of generators operating in the day ahead and intraday market (managed by OMIE) and the other to the costs or savings associated with ancillary services (managed by REE, the system operator). The analysis in this paper focuses on the adjustment cost, or DC, related to the markets managed by OMIE. The authors recommend further research on the adjustment costs related to ancillary service costs.

[15] Congestion rents are not usually used to reduce the energy component of electricity retail prices and thus do not usually have an immediate impact on consumer prices. They may be used for many other purposes that typically affect future system costs. The IE is unusual in using these rents effectively to reduce current effective retail prices.

- The higher the price of gas in MIBGAS;
- The greater the volume of fossil-fired (mainly gas-fired) generation and the greater the fossil share of total generation;
- The lower the reference gas price;
- The lower the congestion payment to Spain (Iberia) related to the Spain-France interconnector; and
- The lower the share of demand that must pay the compensation (DC).

In granting Spain and Portugal an exception to its general policies with respect to interventions in power market prices, the European Commission required that the Iberian wholesale prices would continue to be accessible to other European Union members. This meant that France could continue to import power at Iberian market prices without paying the demand contribution required of affected Iberian consumers. Although France benefited from the lower Spanish wholesale price, it shared the congestion rent 50/50 with Spain, which had agreed under the IE to share it with all consumers with a DC requirement, including those from Spain, Portugal and Morocco.

The IE reduction in market prices affected infra-marginal generation, i.e., plants with lower marginal costs and offer prices than plants setting market prices. There are two categories of inframarginal plants: merchant plants and regulated renewables generators that sell into the market and have a guaranteed minimum return. The authors consider the merchant plants to be especially relevant[16]. They include nuclear, hydroelectric, and merchant wind and solar plants, all of which – at the time the IE was introduced – were already subject to a €67/MWh price cap introduced in September by Royal Decree-Law 17/2021. That cap applied to 90% of their output, thus leaving 10% that might be sold at market prices if not already contracted at lower prices (which they often were). That cap also did not apply to generators of less than 10 MW or to extra-peninsular Iberia. The system savings obtained by the application of lower prices to inframarginal generation could be used to reduce the costs of the Spanish electricity system, for instance reducing access tariffs and consumer tariffs. However, the savings on inframarginal generation resulting from the IE were small and less than would have been the case if the government had taxed extraordinary profits based on the higher wholesale market price that would have obtained without the IE. Although this paper analyses the effect of the IE on inframarginal generation, it is a topic the requires further research.

Because of its design, the IE certainly did reduce wholesale power market prices relative to what they would otherwise have been, and those wholesale price reductions were reflected in the invoices of those consumers whose prices were indexed to wholesale market prices. However, as just noted, those consumers were also required by the IE to make a demand contribution to subsidize fossil generators; that contribution offset the IE's reduction in wholesale power market prices. Consumers with fixed price contracts were initially exempt from contributing to the financing of the fossil generators. But the Spanish Government legislation effectively required most consumers to finance the generation subsidy when their fixed price contracts were renewed at the end of one year, even when those contracts were multi-year contracts.

---

[16] The regulated renewables are discussed later in the paper, beginning in Section 3.1. They also sell into the market but, ultimately, their returns are regulated so that they earn at least their guaranteed minimum return. The authors are uncertain whether some older renewables, notably wind generation, were no longer subject to regulation and were therefore affected by the IE in the same way as other merchant plants.

In addition to predictable concerns related to government intervention (including concerns about distorted price signals that could lead to operating and investment inefficiencies in Iberia), the most fundamental consequence of the lower MIBEL prices was an increase in demand within Iberia and for exports to neighboring countries, notably France. Indeed, the data available on trade between France and Spain suggests that the IE's implementation immediately led to a substantial increase in Spanish exports to France, resulting in higher levels of Iberian generation and consequently higher costs per MWh to be recovered from Iberian consumers exposed to the wholesale market than would have been the case without the IE. These higher marginal generation costs are additional to the GC and the DC[17] that were both calculated assuming that gas was purchased at MIBGAS prices and that plants had an efficiency of 55%.

## 3. A Conceptual Overview of the Effects of the Iberian Exception

The standard economic framework for analysing a universal subsidy illustrates that it ultimately raises the cost and price of the subsidized product[18]. It is understandable that governments wish to protect consumers against dramatic increases in electricity prices. However, a subsidy acts like a price cap that reduces incentives to cut back on the proportion of energy consumption that consumers would have been willing and able to pay if they faced higher prices. The result is that government pays an even higher price for the energy than had there been no intervention since higher demand increases the market price of electricity. Ultimately, the cost to society rises. The authors accept this standard framework, which applies in the case of the IE.

However, the effects of the Iberian Exception are more complicated for several reasons. The first is that the price limit is introduced through a subsidy paid to fossil-fired generators that set the marginal prices on the wholesale market. It is not a direct subsidy to retail consumers or a cap on retail prices. Second, it is not universal; it applies to consumers whose retail prices are indexed to the wholesale market. Third, these same consumers fund the generator subsidy through a demand contribution. Fourth, that contribution depends on multiple factors, including the level of congestion rents in trade with France and the percentage of demand paying the compensation. As a result, it is not immediately evident whether retail prices are falling or rising due to this policy. The Spanish Government maintains that retail prices are falling significantly because of the IE. To the extent that consumers think this is true, they may well behave in line with the standard economic framework, with all its implications. This certainly applies for exports to France. But, what if the actual effect of the IE is to raise retail prices to Iberian consumers compared to what they would otherwise be? Consumers may be consuming more than they should because they believe prices are lower, even when they are not.

---

[17] Assuming the same amount of congestion rents.
[18] Perkins, J. and Rainaut, C. (2023), The Simple Economics of Energy Prices, https://www.compasslexecon.com/the-analysis/the-simple-economics-of-energy-prices/02-22-2023/

In a complicated situation like this, the only way to assess whether retail prices for Spanish consumers are falling is to compare them to a counterfactual. This section offers a conceptual overview of the effects of the Iberian Exception assuming a counterfactual where there is demand elasticity, at least for some consumers. The overview identifies five effects of the IE, assuming demand elasticity.

1. The privileged fossil generators that would have run in the absence of the IE obtain additional margins under the IE because market prices are higher than they would have been without considering demand elasticity. This is due to a rise in demand, notably from France, resulting from the IE's depressing effect on Iberian wholesale prices, the consequent need to rely on additional and more expensive generation to meet that demand, and the increase in market clearing prices reflecting the higher cost of additional generators.

2. The additional privileged fossil generators that operate because of the IE, and that would not have operated otherwise, obtain revenues and margins.

3. The non-privileged generators (merchant and regulated) see their revenues decline because wholesale market prices are lower, although some regulated generators may recover revenues in future.

4. The system experiences a reduction in the margins (between market prices and price caps on inframarginal plants) that could be used to reduce system costs and consumer prices, again because of a lower wholesale market price under the IE.

5. The immediate impact on the affected consumers (i.e. whose prices are indexed to the wholesale spot price) depends on the difference between the reduction in wholesale prices and the size of their demand contribution (DC) to finance the subsidy (GC) to fossil generators; and this difference depends inter alia on demand elasticity.

### 3.1 Impact on generators

In thinking about the actual Iberian power prices and volumes in the first 100 days of the IE relative to the power prices and volumes that might have prevailed in a counterfactual, non-IE world, it is perhaps easiest first to consider supply and demand curves Scf and Dcf as they would have been in the absence of the imposition of the IE, which the authors refer to here as the counterfactual (cf). As shown in the graphic below, Figure 3, the pre-IE supply and demand curves Scf and Dcf result in the counterfactual (cf) equilibrium market prices and quantities Pcf and Qcf at the intersection of the Scf and Dcf supply and demand curves. The supply curve rises to reflect the rising cost of generation to meet higher levels of demand. The declining demand curve reflects demand elasticity with respect to price.

**Figure 3.** Illustrative Iberian Supply and Demand Curves and Resultant Power Quantity and Price in a Counterfactual scenario without the IE

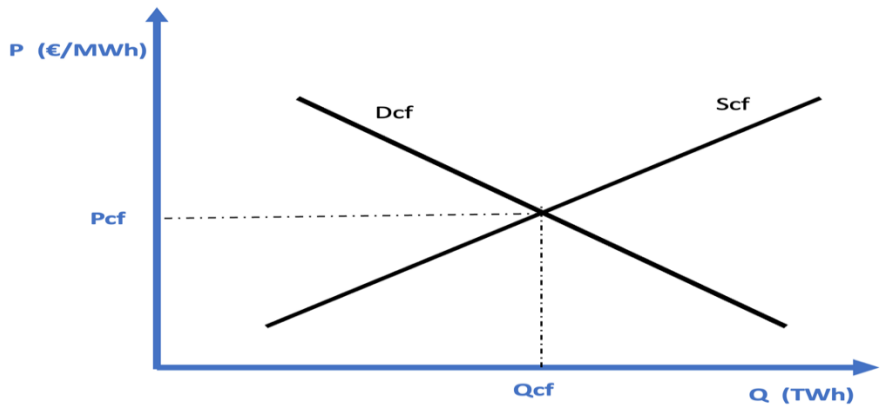

As discussed in Section 2, the IE complicated the determination of supply and demand balances by subsidising the cost of gas for the fossil generators through the generation contribution (GC). This reduced their offers of generation supply by the amount of the GC, as is depicted by the lower (dashed) supply curve Sact in Figure 4 below; this reflects the situation that existed under the IE. However, for Iberian consumers whose power prices were linked to wholesale market prices, the IE also required those affected consumers to pay a demand contribution (DC) that helped finance the GC and that substantially offset the IE's reduction in wholesale power prices. This is reflected in the higher (dashed) demand curve Dact in Figure 4.

As an exception to the demand contribution that the IE imposed on Iberian consumers with prices linked to the wholesale market, one important component of demand, exports to the French power market, did not pay the DC charge. The minimum[19] for France of the French-Iberian power price differentials were half of the congestion rent on that interconnector; with Spain obtaining the other half (which it agreed to share with all consumers subject to the DC). Together, the French demand and the demand of large consumers were elastic with respect to price.

---

[19] There were likely other benefits not captured in the analysis of congestion rents. For instance, the Spanish regulator opened an investigation into whether Spanish retailers who bought electricity in the Spanish intraday wholesale market for sale into the French market had colluded. Whether they had colluded is unclear and is contested by the companies being investigated, but this is evidence of the potential for lower Iberian market prices to have further reduced costs in France.

**Figure 4.** Illustrative Iberian Supply and Demand Curves and Resultant Power Quantity and Price with (act) and without (cf) the imposition of the IE.

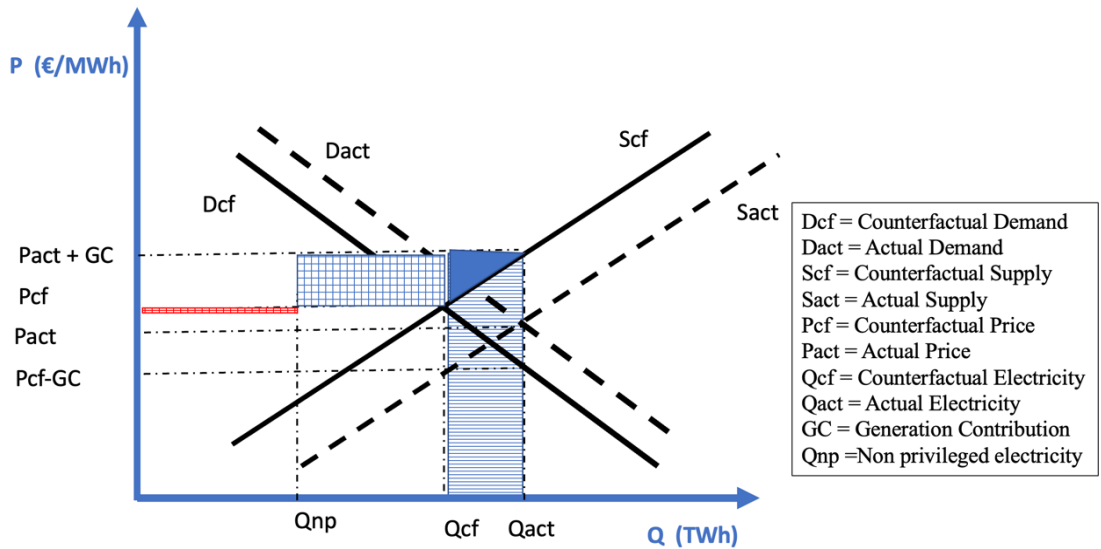

If (as the Ministry apparently assumes) the IE's reduction in generators' supply offers had had no effect on the amount of power consumed, that is, if the quantity demanded under the IE had remained at the counterfactual level Qcf, then the market clearing price would have declined by the full amount of GC, from Pcf to Pcf-GC, where the Sact supply curve corresponds to pre-IE demand quantity Qcf. However, given that demand related to the Iberian market, notably the demand stemming from France, was sensitive to prices (as reflected in the downward-sloping demand curve D), demand under the IE increased relative to what would have been pre-IE levels, leading to new equilibrium points Qact and Pact at the intersection of the IE-related supply and demand curves, Sact and Dact respectively. This shows that the likely IE-related reduction in wholesale power prices relative to the counterfactual, Pcf - Pact, was less than GC, contrary to what the authors believe was a Ministry assumption in its assessment of the benefits of the IE.

The implications of the foregoing overview for generators can perhaps be conveyed most simply by focusing on the shaded sections of the graphic.

**First, the blue hatched box** (between prices Pcf and Pact + GT, for quantity Qnp to Qcf) corresponds to an increase in compensation for the privileged generators that were operating in merit order before the IE was introduced and that were also operating under the IE. Market-clearing prices for all volumes of generation under the IE decreased from Pcf to Pact. However, for privileged generators receiving the GC subsidy—which is what financed the reduction in their offer prices to the market operator OMIE, which then resulted in the downward shift of the supply curve from Scf to Sact, and which further caused the decrease in market-clearing prices to Pact—their total compensation per MWh increased from Pcf in the absence of the IE to Pact + GC with the introduction of the IE. This increase in the compensation per MWh of privileged generators is a direct consequence of two factors: the higher demand and generation volumes brought about by the IE relative to what would have prevailed in the absence of the IE, and the upward slope of the supply curve. The simple fact behind the supply curve is that the costs and offer prices of the additional units of generation brought on to meet the higher demand volumes under the IE would be higher than those of the marginal units that would have been dispatched without the IE. As shown by the intersections of the Sact supply curve with the Dact volumes, the marginal offer prices corresponding to the demand Qcf would increase from Pcf – GC to Pact. Those generators continued to supply the same volumes using the same gas supplies, producing power with the same conversion efficiencies / heat rates, and paying the same carbon charges, but the extent to which their costs changed is an empirical question that deserves further research[20]. To the extent that their costs were the same, these generators earned additional profit margins from the IE.

**Second, the dark blue and light blue column** in Figure 4 corresponds to additional generation that only runs due to the increased demand under the IE. For the privileged generators supplying the incremental volumes Qact – Qcf, the entire column bounded by those volumes and the Pact + GC price represents increased revenues and therefore increased costs passed on to consumers buying in the spot market. The minimum additional profit margins of the IE for generators supplying the additional generation are reflected in the triangle at the top of the column (in dark blue in Figure 4). The lowest-cost units first dispatched would have a minimum profit margin determined by the difference between the market-clearing price of Pact + GC and their costs, as shown in the Scf supply curve. The last units dispatched would have supply costs equal to the market-clearing price and no incremental profits. However, these are minimum additional margins; some industry experts would argue that most generators tend to include a profit or "industrial" margin of 5% to 10% on the total amount of the power on offer. If so, the increased revenues brought about by the IE would have meant substantial increases in net profits for these generators.

**Third, the thin red horizontal area (below Pcf)** in Figure 4 reflects the potential losses of some non-privileged inframarginal generators compared to what would have happened without the IE. The wholesale price reduction resulting from the IE (from Pcf to Pact) reduced their market revenues (whose output is represented in Figure 4 by Qnp). As discussed in the previous section, the negative effect on the inframarginal merchant generators was limited. Most (90%) of inframarginal merchant plant output already faced a price cap of €67/MWh, so the net impact of the reduced wholesale market price affected only 10% of the incremental revenue (the red horizontal area beneath the counterfactual price Pcf)[21]. Furthermore, for other inframarginal plants, the regulated renewables and cogeneration, the adverse consequences may have been limited by guaranteed minimum rates of return (to be recovered later)[22]. The authors of this study think that even market prices depressed by the IE would still have provided returns above the minimum guarantees and thus would require no makeup payments after the 100 days addressed in this paper. As noted in Section 4, NERA has a different view on this matter and this topic may deserve further research.

---

[20] Note that gas costs under the counterfactual for the privileged generators may not be the same as under the IE if generators were purchasing a fraction of their gas at international spot prices rather than on fixed price contracts.

[21] To clarify the point with an example: Suppose that the application of IE reduced market prices from Pcf = €200 to Pact = €150. Non-privileged generators (Qnp) subject to price caps would be receiving €67/MWh plus 10% of the difference up to the market price (see Section 2). Income would go from receiving €80.3 (67 +0.1*(200-67)) to €75.3 (67 +0.1*(150-67)), reducing their income by €5/MWh, which is represented by the red area in Figure 4.

[22] The authors are uncertain whether some older regulated generation, notably wind power stations, should be treated as merchant generation because they were no longer subject to regulation. This deserves further research.

**Fourth, the IE's price reduction for generation amounts up to Qcf (the area between Pcf and Pact for generators not otherwise price-capped)** reflects revenues that otherwise could potentially have been taxed to finance system costs and to benefit consumers through lower access tariffs or reduced charges to cover policy costs[23].

### 3.2 Impact of the IE on consumers

**Fifth, to illustrate the impact on consumers**, this subsection compares two approaches to analyzing the counterfactual: one with and one without demand elasticity. To simplify the comparison, the analysis ignores congestion rents, which help to finance the generator compensation (GC). This is to explain how demand elasticity can change the demand contribution (DC), which is the other main revenue source funding the GC.

Figure 5 shows one illustrative case concerning possible differences in consumer power costs per MWh under the IE (Pact + DC) and in a counterfactual case without the IE (Pcf). The left-hand side of Figure 5 represents the Spanish Government's approach, which assumes a vertical demand curve, so there is no change in demand when prices rise or fall. In that case, the consumer benefits are represented by the blue hatched area. The counterfactual wholesale price (Pcf) equals the market price under the IE (Pact + GC). That counterfactual price is greater than the wholesale price the consumer pays under the IE (Pact + DC). In other words, GC is greater than DC when there is no demand elasticity; customers always benefit.

By contrast, the right-hand side of Figure 5 represents an approach with a downward sloping curve that reflects demand elasticity. In this counterfactual, due to demand elasticity, wholesale prices (Pcf) are higher and demand lower than when following the Ministry methodology. For instance, in Figure 5, the red hatched area reflects that, with demand elasticity, the counterfactual price (Pcf[24]) paid by the consumer could be less than the price paid under the IE (Pact+DC). In other words, consumers may pay more under the IE than had the IE policy not been introduced.

---

[23] In the same sense as in note 21, system revenue from €119.7/MWh (0.9*(200-67)) to 74.7 (0.9*(150-67)). In other words, the €50/MWh price reduction because of the EI is divided between a reduction in income from non-privileged generators (€5/MWh) and a reduction in system income (€45/MWh).

[24] Note that Pcf will be less than Pact + GC whenever the demand curve is decreasing.

**Figure 5.** Illustrative Iberian Supply and Demand Curves and Resultant Power Quantity and Price with (act) and without (cf) the IE. Ministry (left) and alternative (right) counterfactual. Impact on consumers.

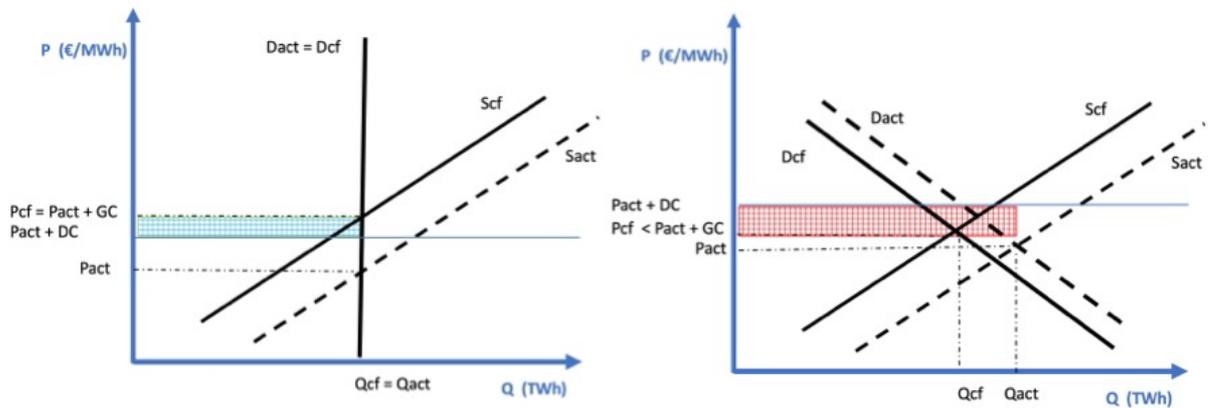

| | |
|---|---|
| Dcf = Counterfactual Demand | Pact = Actual Price |
| Dact = Actual Demand | Qcf = Counterfactual Electricity |
| Scf = Counterfactual Supply | Qact = Actual Electricity |
| Sact = Actual Supply | DC = Demand Contribution |
| Pcf = Counterfactual Price | GC = Generation Compensation |

As stated, this is simply an illustrative case. A final determination of the relative sizes of the GC and the DC per MWh would need to consider two factors that enter the determination of the total amount of the GC and the amount of the DC per MWh. These are: 1) the amount of generation supplied by privileged generators; and 2) the fraction of total Iberian power consumption that is used by affected customers. This is because, setting aside the contribution of congestion rents to the financing of the total GC, 1) the total GC is determined essentially by the governmentally determined GC per MWh of the privileged generation multiplied by the amount of that privileged generation and 2) the DC (which is a levy per unit of consumption) is determined essentially by dividing the total GC by the consumption of affected consumers. The higher the share of total Iberian power consumption that is supplied by privileged generators, the higher the total DC. The higher the share of Iberian consumption by affected consumers, the lower the DC on a MWh basis.

## 4. The Ministry's Methodology for Estimating the Effects of the Iberian Exception on Spanish Consumers

It is difficult to simulate what would have been the market dynamics and the effects of the Iberian wholesale power market if the IE mechanism had not been introduced, i.e., to develop a counterfactual scenario. That said, several analysts, notably the Ministry, have relied on what the authors of this paper view as overly simplified assumptions relating to counterfactuals. This section focuses on the authors' understanding of the methodology used by the Ministry.

## 4.1. Description of the Ministry methodology and analysis

In order to assess the effects of the IE on Spanish consumers during its first 100 days, the Ministry focuses on a comparison of: 1) the overall power costs under the IE of Spanish consumers affected by wholesale power prices, that is, the actual power market prices received by privileged fossil generators and passed on to affected consumers, plus the actual demand contribution (DC) also paid by affected Spanish consumers; and 2) a projection of the power market prices that would have been passed on to affected consumers in the absence of the IE. To make a projection of power market prices in the absence of the IE, the Ministry approach necessarily develops a set of assumptions about a counterfactual scenario that would have occurred in the absence of the IE.

To estimate wholesale market prices in its counterfactual, the Ministry methodology begins with the actual MIBEL hourly data on market prices for the first hundred days of the IE that are published by the Iberian market operator (OMIE)[25]. It then increases those prices by the hourly compensation (GC) amount that privileged generators received under the IE, also as recorded by OMIE. This results in estimated wholesale power market prices in the counterfactual that are markedly higher than the actual prices under the IE.

The Ministry then estimates the average hourly savings of consumers affected by the IE as the difference between its counterfactual wholesale market price (with no demand contribution), which is 2) above, and the total of the actual wholesale price under the IE and the actual demand contribution (DC) required of affected consumers under the IE, which is 1) above.

The Ministry may have considered possible changes in domestic and export demand due to the different effective costs of power for consumers affected by the IE (market prices plus demand contributions) relative to its estimation of costs (simply its counterfactual market prices) in its counterfactual. However, to the knowledge of the authors, there is no official public document or other evidence to confirm that the Ministry's calculations of the effect of the IE took any such potential differences in demand into account. The Ministry methodology described in this paper assumes that they did not quantify any potential demand-side effects.

Figure 6 below depicts the three key patterns of prices per MWh for affected consumers over the first 100 days of the IE according to the Ministry methodology. The gray curve traces actual wholesale market prices under IE. The blue curve reflects what affected Iberian consumers actually paid under the IE; this includes the actual wholesale price and the demand contribution (DC) made by Iberian consumers as part of the GC subsidy paid to fossil generators. The red curve is the Ministry's estimate of the counterfactual wholesale prices had the IE mechanism not been applied. This red curve is simply the pattern of actual market prices plus the generator compensation (GC) received by the

---

[25] OMI-Polo Español S.A. (OMIE) was appointed in 2015 as Nominated Electricity Market Operator by the competent Spanish and Portuguese authorities.

.

privileged generators. The red curve is usually above the blue curve, implying significant savings from the IE for affected consumers.

**Figure 6.** MIBEL hourly market prices (grey), hourly market prices plus the demand contribution (blue) and wholesale prices (r) had the IE not been introduced, according to the Ministry (15/06/22 – 23/09/22).

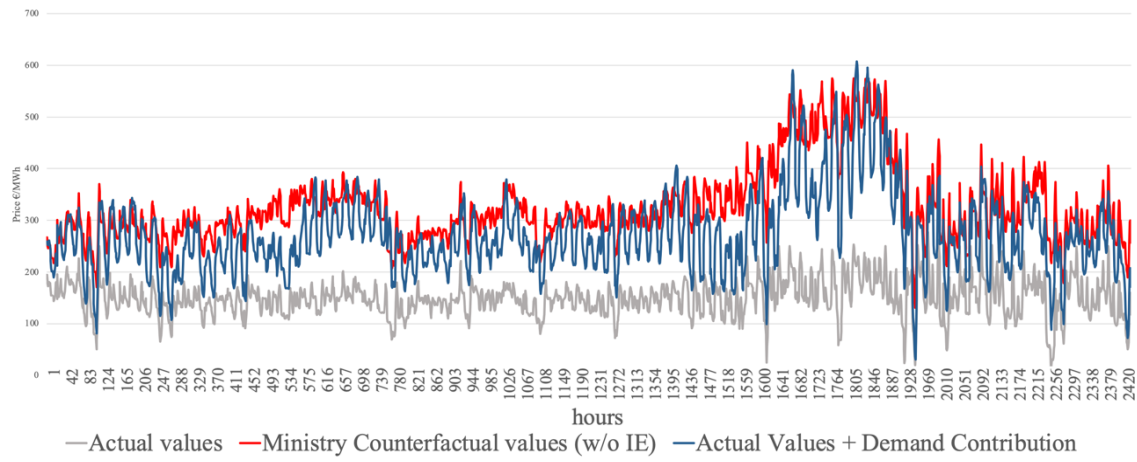

Table 2 quantifies the results presented in Figure 6. According to this interpretation of the Ministry's counterfactual, if the IE had not been introduced, the average wholesale market price would have been about 120% higher than the actual price, assuming the same demand, but there would have been no demand contribution required of affected consumers. This suggests that, relative to the Ministry's counterfactual, the IE yielded savings for Spanish consumers of 20%, or about 1,400 million euros (53.9 x 26,022 = 1,402,586), over its first 100 days. As noted earlier, these results ignore the impact of lower net consumer power costs under the IE on the quantity of power demanded (i.e., the results ignore any possible demand elasticity with respect to price).

|  | Actual values (under IE) | Ministry Counterfactual values (w/o IE) | Ministry Counterfactual values vs. actual values (%) | Ministry Counterfactual values vs. actual values |
|---|---|---|---|---|
| **Wholesale average MIBEL price (€/MWh)** | 148.5 | 326.8 | 120 | 178.3 |
| **Average demand contribution for affected Iberian consumers (€/MWh)** | 124.4 | 0 | - | -124.4 |
| **Average total cost for affected Iberian consumers (€/MWh)** | 273 | 326.8 | 20 | 53.9 |
| **Average hourly total Spanish demand (MWh)** | 26,022 | 26,022 | 0 | 0.00 |

**Table 2.** Comparison of the actual values under the IE and the Ministry counterfactual for: average wholesale electricity market prices, the demand contribution made by affected consumers, the total cost for the contributing consumers, and total hourly Spanish demand (15/06/22 – 23/09/22).

Note that the first three rows of the table refer to Iberia as a whole. Since the Ministry analysis focuses on the effect of the overall Iberian costs on only the Spanish subset of customers, the last row breaks out total demand volumes for Spain alone.

Under the assumptions apparently used by the Ministry in its counterfactual, demand would remain unchanged for Portuguese and export customers. Consequently, total Iberian generation volumes, and thus generation costs per MWh, would also remain unchanged from volumes and costs under the IE.

### 4.2 Commentary on the Ministry analysis

That the Ministry calculations fail to reflect possible demand elasticity for both Iberian consumption and exports to France and Morocco is, in the view of the authors of this paper, a flaw in their analysis. In the absence of the IE, MIBEL wholesale market prices would have been substantially higher and demand, especially export demand from France, substantially lower, thereby reducing generation volumes and the marginal cost of generation in Iberia. As does this paper, NERA Economic Consulting (NERA), in a series of articles by Arnedillo et al. (2023)[26], also identify increased exports to France as a significant change resulting from the IE. The following section discusses and quantifies the relevance of the Ministry´s oversight.

NERA also charges the Ministry with overestimating the consumer benefits of the IE by including the IE's reduction in the revenues of renewables generators as a consumer benefit but ignoring the guaranteed minimum returns in several renewables generation contracts that could cause a later reimbursement and increase in consumer costs. While this effect is possible, the authors of this paper do not think as a practical matter that the IE reduced renewables' returns below guaranteed levels. If that is correct, a reduction during the first 100 days of the IE would not lead to make-up consumer charges at some future point. There is, however, some uncertainty about whether some old renewables (especially wind power) were no longer subject to regulated returns and therefore merchant plant with the rights and limits (€67/MWh) that implied; in that case, they would have lost potential margins under the IE. However, the issue of the effect of the IE on renewables may deserve further research.

---

[26] See three articles by Arnedillo et al. in *El Periódico de la Energía, "Análisis de los efectos de la excepción ibérica"* (1) (2) and (3), March 2023.

## 5. This Paper's Analysis of the Effects of the IE on Spanish Consumers

This section presents an alternative quantitative analysis of the effects of the IE on affected Spanish consumers during its first 100 days.

Any counterfactual relating to the IE must, in the view of the authors, examine the potential impact of the IE on both supply and on demand. As discussed in Section 4, the Ministry's counterfactual focuses on the supply side of the market. Lowering wholesale prices was a direct and obvious target of the IE's subsidies to the generators that determine wholesale power market prices, but this effect was substantially offset for affected Iberian consumers by the IE's demand contribution requirement. However, French imports were not subject to this requirement and the IE's reduction in market prices relative to what would have prevailed in its absence clearly increased the attractiveness of French imports from Spain, as this paper demonstrates in Section 5.2 below. Some analysts[27], notably including NERA but not the Ministry, mention such demand effects. This paper presents an analysis of the IE's consequences for Spanish consumers that quantifies those effects on demand.

This section begins with a qualitative overview of two alternative counterfactuals that emphasize the impact of demand elasticity. These two counterfactual scenarios are then, in succeeding subsections, quantified and used to assess the effects of the IE.

- The first scenario reflects demand elasticity in Spain; and
- The second scenario reflects demand elasticity in Spain as well as in export markets--specifically potential changes in electricity trade with France. This is the counterfactual used to assess the impact of the IE on Spanish consumers and other affected parties.

---

[27] Other authors, such as Brito (Brito, 2022) recognize that demand can be elastic, but do not estimate the effects of that elasticity. Hidalgo et al. (Hidalgo et al., 2022) use Bayesian structural models to build a counterfactual, incorporating causal relationships, projecting the time series of energy prices for PVPC customers, and thus estimating savings. However, they study the effect on only a limited part of the market, and they do not incorporate demand elasticity in their modeling.

Both new counterfactual scenarios utilize OMIE hourly data on actual supply and demand offers under the IE to estimate what generators and consumers (or their representatives) would have bid in the absence of the IE.

- The market clearing price projected in these counterfactuals reflects the highest generator offer price needed to equate supply and demand volumes.
- Gas- and coal-fired generators submit offer prices that equal the offer price they submitted under IE, plus the extra GC compensation paid to those generators under the IE.
- Other producers, especially dispatchable hydro and some intermittent renewables, whose offer prices reflect their opportunity cost, submit offer prices in line with gas-fired generation.
- Some renewable and other generators are subject to price caps (€67/MWh) that continue to apply in the assumed absence of the IE.

This paper estimates a counterfactual supply curve for the Iberian market—that is, the generation offers that would have prevailed in the absence of the IE—using the actual hourly offers made by generators in the first 100 days of the IE, plus the actual generation contribution (GC) received by privileged generators. This fact-based approach seems reasonable and logical to the authors of this paper and is a quantification of the conceptual Scf supply line depicted in Figure 4, which was presented earlier. This approach parallels that used by the Ministry, so differences in the conclusions of the Ministry and this paper do not arise from the determination of a counterfactual supply curve.

This paper also uses actual (IE) data on Iberian demand offers, after adding the actual demand contribution (DC), as a reasonable and logical basis for modelling the counterfactual demand curve, depicted as Dcf in Figure 4.

As discussed earlier in Sub-section 3.1, the Ministry, to the best of the authors' knowledge, made no explicit determination of a demand curve and appears implicitly to have assumed a completely inelastic (vertical) demand curve, i.e., a volume of demand unchanged by price. The Ministry's omission of demand considerations in its analysis almost certainly is the major source of the disparity in the conclusions of the Ministry and this paper with respect to the effects of the IE on Iberian power consumers.

The authors do not have access to comparable time series of actual supply and demand offers readily available for the French market; they have only the actual hourly market prices that applied to trade over the interconnector in the first 100 days of the IE. Because of the availability of power at Iberian market prices depressed by the IE, actual French market prices may have been lower than they would have been in the absence of the IE. However, the quantified effect of the IE on French power market prices cannot be determined with precision and the authors argue later in this section that the IE probably had little effect on French wholesale prices.

With the foregoing caveat, because the IE directly caused Iberian wholesale power market prices to be lower than they would have been otherwise, the total demand for Iberian electricity (over the Spanish interconnector with France and from Iberian industrial and commercial consumers) and total Iberian generation volumes were higher under the IE than they would have been in its absence. Because of this higher generation volume, marginal Iberian generation costs under the IE were also higher than they would otherwise have been, and these differences are reflected in the counterfactuals studied below.

**5.1 First alternative counterfactual reflecting elasticity of demand in Iberia**

This paper's first alternative counterfactual focuses on Iberia. The first part of this section is conceptual and explores the qualitative impact of the IE on Iberian electricity demand, the Iberian electricity market equilibrium and affected consumers in Spain. The second part presents quantitative results.

5.1.1 Conceptual overview

The authors' first counterfactual recognises that residential demand is generally quite inelastic in the short run, but also that some industrial and commercial demand is elastic and will respond to day-ahead spot prices. In practice, large consumers and their suppliers enter demand offers at prices that enable producers to make and sell their products and services at a profit[28].

Under the Iberian Exception (the actual situation), by reference to the day ahead MIBGAS market, large consumers could know the approximate level of the demand contribution (DC) that they would have to pay in advance of making their bids into the day ahead electricity market. For most large consumers, their bids for power supply could thus reflect the net cost to them after taking account of (i.e., adding the amount of) their demand contributions. They could then enter bids for energy at prices up to the value of that energy to them.

To reflect supply, demand, and prices under the IE, this paper uses the actual data for the IE's first 100 days. Graphically, in combination with the demand of smaller consumers, the demand curve of large consumers (Dact in Figure 4) intersects with the supply curve (Sact in Figure 4) established by the offers of generators, resulting in the actual market-clearing prices and volumes shown in the OMIE records for the first 100 days of the IE. The Ministry use these same volume numbers as the basis for its counterfactual analysis. In so doing, they implicitly assume that the demand curve is vertical. To the contrary, the analysis in this paper assumes that the demand curve is downward sloping.

The modeling of this paper's first counterfactual reflecting Iberian demand elasticity leads to different wholesale equilibrium prices and volumes than the Ministry´s approach. This new counterfactual equilibrium reflects the absence of generator subsidies, a consequent need for generators to look entirely to the wholesale market to cover their costs, considerably higher offer and wholesale spot prices, the absence of consumer demand contributions, higher net power costs per MWh for consumers and—reflecting the price

---

[28] The hourly energy demand of industrial and large commercial consumers relative to price is related to their production processes and the market value of their products. Their offer prices to buy electricity in the market reflect a limit above which they would have no incentive to buy more power and produce a higher volume of products or services.

elasticity of large consumers—different volumes of demand and generation (Dcf and Scf in Figure 4).

The calculation of the IE's net benefits for affected consumers depends on the difference between what consumers would have paid in the counterfactual (Pcf, reflecting wholesale prices with no IE-related charges) and what they actually paid under the IE (a lower wholesale price Pact plus the IE-related demand contributions). Provided that price-elastic consumers recognize that their net power prices in the counterfactual are higher than what they paid under the IE (including their demand contribution to the payment of generator subsidies), their demand (and thus Iberian generation) would have been lower in the counterfactual. The essential difference with the Ministry's calculation is that this paper's counterfactual equilibrium depends both on supply and demand (in this first counterfactual case, lower Iberian demand) and the effects of lower demand on generation supply costs.

The analysis in this counterfactual follows the Ministry's approach to congestion rents, namely, to include the benefits of those rents under the IE and ignore them in the counterfactual. This is justified on the grounds that congestion rents are not normally used to reduce the energy component of retail prices. Ignoring them in this counterfactual increases the estimated benefits of the IE, but the difference is very small, as illustrated in Section 5.2 which examines a counterfactual that includes trade with France.

5.1.2 Modeling process and results

The quantitative modeling of this paper's first counterfactual uses publicly available hourly data from OMIE for supply and demand offers, as well as OMIE's data on the demand contribution (DC), generation compensation (GC) and export volumes under the IE.

Assuming no change in international trade compared to what occurred, the modeling estimates the effect of the IE on Iberian demand by large consumers. Using the same format as Figure 6 (but different colours for the counterfactual), Figure 7 compares the actual market prices (gray curve), those same market prices plus the demand contribution (blue curve), and estimated wholesale prices had the IE not been applied (green curve). As in the government's counterfactual (their red curve) the green curve in this alternative counterfactual remains mostly above the blue curve, reflecting savings experienced by the affected Iberian consumers.

**Figure 7.** Hourly market prices (grey), hourly market prices plus demand contribution reflecting actual results (blue) and prices that the market would have reached (green) had the IE not been in force. This is the first counterfactual assuming demand elasticity only in Iberia itself. (15/06/22 – 23/09/22) -2400 hours.

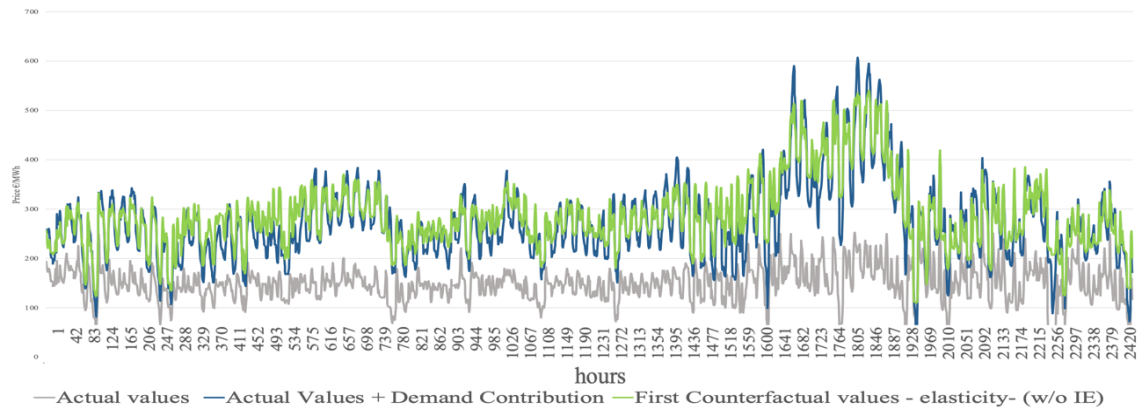

Table 3 compares the actual values resulting from the Iberian Exception with the counterfactual values of the scenario summarized in Figure 7. Absent the IE, given the demand elasticity reflected in OMIE's data on actual consumer purchase bid prices and the generator supply curve reflected in actual OMIE generator offer prices, Spanish demand in this counterfactual would have been lower by 3%, wholesale prices would have been higher by 161 €/MWh, but the IE's demand charge (i.e. 124 €/MWh) would not exist, so the net cost for Iberian consumers would have been higher by 36 €/MWh, or 13%. Although the IE savings of 13% are lower than the savings estimated using the Ministry's methodology, this first-pass analysis tends to support the Ministry conclusion that affected Spanish consumers (i.e., PVPC consumers and larger consumers whose prices are indexed to the spot market) would have benefited from the Iberian Exception. However, this estimate almost certainly overstates the consumer benefits of the IE because it assumes that only large Iberian consumers would have reduced demand in response to significantly higher wholesale prices in the counterfactual scenario. If more consumers had reduced demand in the counterfactual, wholesale prices would have been lower than shown in Table 3 and the benefits of the IE less.

| | Actual values (under IE) | First Counterfactual values (w/o IE) | First Alternative Counterfactual values vs. actual values (%) | First Alternative Counterfactual values vs. actual values |
|---|---|---|---|---|
| **Wholesale average MIBEL price (€/MWh)** | 148.5 | 309 | 108 | 160.5 |
| **Average demand contribution for affected Iberian consumers (€/MWh)** | 124.4 | 0 | - | -124.4 |
| **Average total cost for affected Iberian consumers (€/MWh)** | 273 | 309 | 13 | 36.1 |
| **Average hourly total Spanish demand (MWh)** | 26,022 | 25,214 | -3 | 808 |

**Table 3.** Comparison of the actual values under the IE and the first alternative non-Ministry counterfactual (reflecting Spanish demand elasticity) for: Iberian average wholesale electricity market prices and the demand contribution made by affected consumers and the total cost for the contributing consumers; and for average Spanish hourly demand (15/06/22 – 23/09/22).

More importantly, this first counterfactual scenario, as apparently does the Ministry analysis, ignores the effect of the IE's reduction in Iberian wholesale market prices on French imports, the increase in Iberian generation required to meet that demand increment, the associated increase in the marginal cost of that additional generation and the consequent increase in power market prices in Iberia. As suggested by the conceptual graphics in Section 3, these supply cost considerations can have a significant effect on net consumer costs per MWh. This last point is addressed quantitatively in the next subsection, which considers the supply-side implications of the higher exports to France that were brought about by the IE.

As a lead-in to the analysis of the effects of French demand on Iberian market prices in this paper's second counterfactual analysis, it is worth commenting briefly here on the justification for using the actual French wholesale prices (during the 100 days studied) as the basis for analyzing flows over the interconnector and for estimating the resulting congestion rents.

- First, the new equilibrium hourly Iberian power market prices determined in the first counterfactual are used to estimate what the congestion rents with France would have been in the absence of the IE, assuming that French market prices are the same in this paper's second counterfactual case as they were under the IE[29].

---

[29] These rents are calculated as the difference between the system market prices in France and Spain in each hour; with the rents divided evenly between the two systems. For both alternative counterfactuals, the congestion rents would have been substantially lower than they were under the IE.

- Second, the assumption of maintaining the same French market prices is open to challenge, but in the view of the authors is reasonable and is unlikely to have a substantial effect on the conclusions. Although the percentage difference in interconnector flows between actual flows and the counterfactual estimate may seem large, this is a large difference in a number that is small relative to the size of the overall French power market[30].

- Third, given major availability problems in the French nuclear sector (and further problems with hydro generation) during the first 100 days of the IE, French power market prices were highly likely to have been based largely on the costs of gas-fired plants, either in France or elsewhere. In that case, absent the IE, French and Iberian market prices would have tended to be roughly similar, thus resulting in no large and continuous price- and demand-driven flows over the interconnector in either direction. This is also suggested by the interconnector flows in both directions prior to the IE that are shown in Figure 8 in Sub-section 4.2 below. Nor would the modest price differentials on flows between the two markets that did take place generate large congestion rents per MWh.

- Fourth, as shown in Table 3, the discount in Iberian wholesale prices caused by the IE is estimated to be enormous, over 50%. Not surprisingly, given any such discount, French imports and exports change radically between actual history and this paper's counterfactual scenario without the IE, with France basically importing to the limit of the available transmission capacity after the implementation of the IE, at substantial market price differentials, thus yielding significant rents, contributions to the GC and reductions in the DC. This argumentation also suggests that rents in this paper's second counterfactual would be substantially lower than were actual rents, as is quantified in the following section.

## 5.2. Second alternative counterfactual reflecting elasticity of Iberian industrial demand and exports to France

Given the significant reduction in Iberian wholesale market prices caused by the IE, it seems unreasonable to ignore, as the Ministry methodology appears to do, the likelihood that the Iberian Exception would affect international trade. Figure 8 shows the hourly exchanges with France for the nine months preceding and the first 100 days following the imposition of the IE. The graphic clearly shows a major change in the pattern starting the day when the IE was introduced; basically, French exports to Spain stopped and exports from Spain increased essentially to the limit of interconnector capacity.

---

[30] While the change in the average volume of energy exchanged on the interconnector is striking in percentage terms, the maximum hourly interconnector flow was less than 2 GWh, which is only 2% of the French market, albeit a higher percentage of the Iberian market. This suggests that exports from Spain were unlikely to significantly affect the French equilibrium market price.

**Figure 8.** Actual hourly flow of the Spain-France interconnection (MWh). September 2021-September 2022.

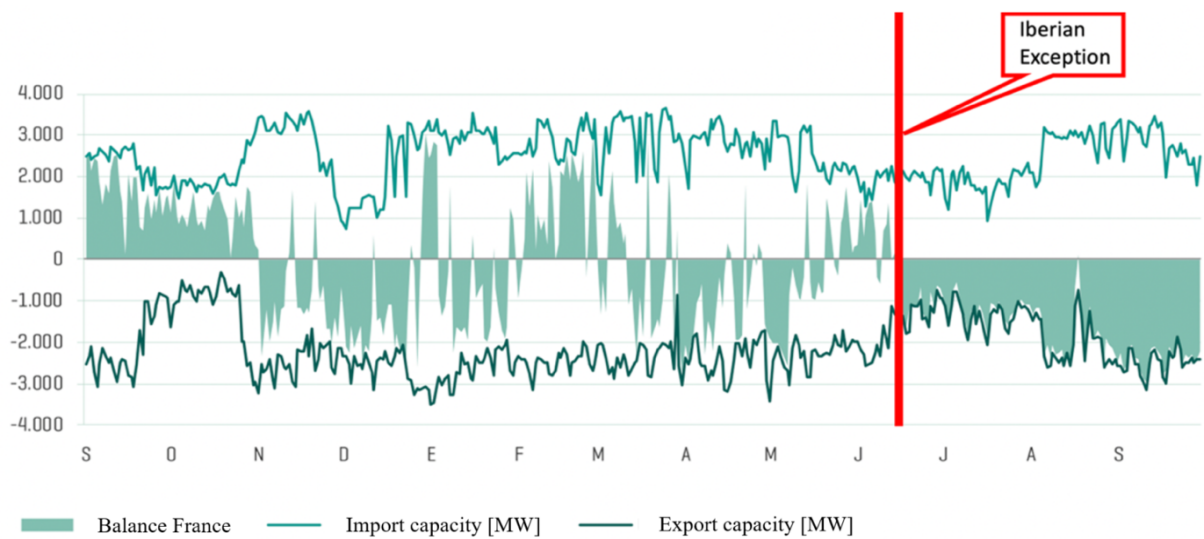

To make quantitative estimates of the implications of this paper's second counterfactual scenario, the Iberian wholesale prices calculated in the first alternative counterfactual scenario are compared with the actual French wholesale prices for each hour over the first 100 days of the IE. For the reasons explained in Section 5.1, in the authors' view, the IE would not have substantially changed French wholesale prices. It is therefore reasonable to use the information available on actual French prices under the IE as the basis for making an initial assessment of the impact of the IE on trade and of the resulting congestion costs. The assumed direction of the flow over the interconnection is always from the lower to the higher priced market in each hour. The price information is used to calculate a congestion rent (per MWh and in total) over the 100 days.

Below, the authors' quantitative model compares the consequences for affected consumers under the IE with those in the second counterfactual, which reflects Iberian and French demand elasticity and the potential for changes in the flows between Spain and France. The direction of flow on the interconnector reflects the price difference between the Iberian and French wholesale markets. The change in behavior reflects the significant reduction in Iberian market prices resulting from the application of the Iberian Exception. If the IE mechanism had not been introduced, i.e., as in this second counterfactual, Iberian wholesale prices would sometimes have been likely to be higher than French prices. Figure 9 captures the model's estimated price differences between the two markets under the Iberian Exception and in this counterfactual scenario during the first one hundred days (2400 hours) of the IE's operation. It is important to note that, as foreshadowed in Section 5.1, the modeling of this counterfactual case explicitly takes into account the effect of this scenario's reduced French demand on Iberian generation volumes and marginal costs relative to the actual generation volumes and costs under the IE.

**Figure 9.** Price differences between Spain and France with and without the application of the IE mechanism during the first 100 days of operation of the IE. From June 15 to September 23, 2022 – 2400 hours.

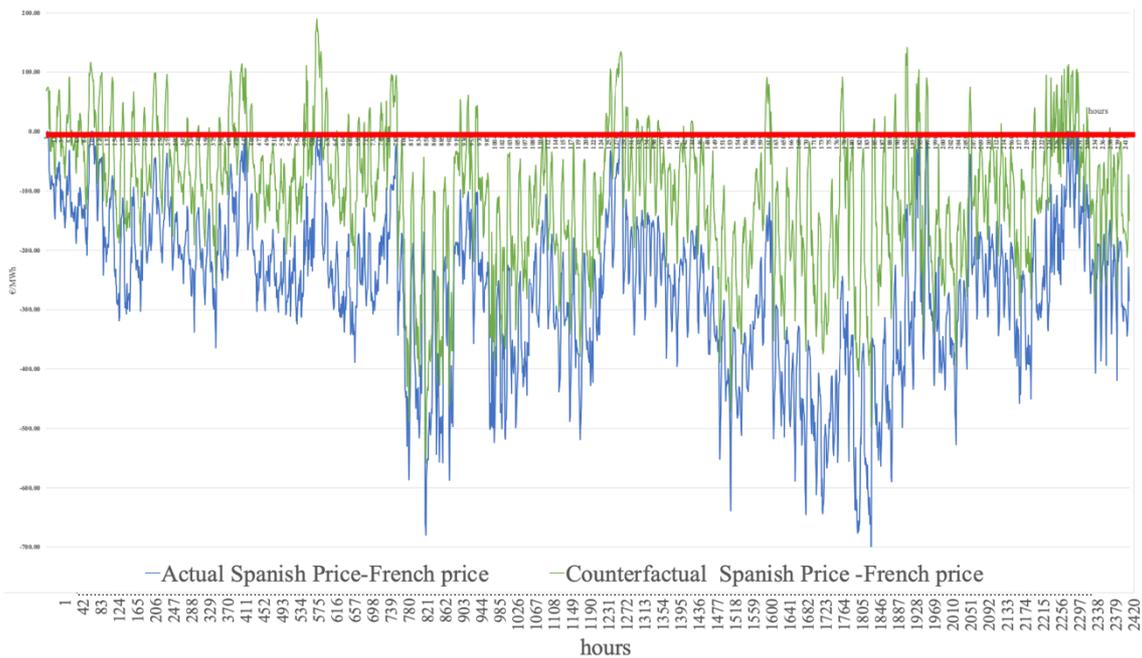

The blue curve represents the difference in French and Spanish wholesale prices under the IE. Under the IE, French prices are systematically higher than in Iberia, which explains the flows in Figure 8, with France always importing from Spain. By contrast, the green curve, which reflects the counterfactual, includes 401 hours when French prices are lower than Spain's, which would have meant that France exported to Spain, reversing the actual flows under the IE.

The counterfactual results obtained in this new simulation are presented in Table 4.

|  | Actual values (under IE) | Second counterfactual values (w/o IE) | Second counterfactual values vs. actual values. | Second counterfactual values vs. actual values |
|---|---|---|---|---|
| **Wholesale average MIBEL price (€/MWh)** | 148.5 | 265.7 | 79% | 117.2 |
| **Average demand contribution for affected Iberian consumers (€/MWh)** | 124.4 | 0 | - | -124.4 |
| **Average total cost for affected Iberian consumers (€/MWh)** | 273 | 265.7 | -3% | -7.3 |
| **Average hourly Spanish demand (MWh)** | 26,022 | 24,496 | -6% | -1,526 |
| **Average hourly Spanish congestion rents (M€)** | 0.251 | 0.064 | -75% | -0.187 |
| **Spanish demand contributors (%)** | 59% | | | |
| **Average hourly Portuguese demand (MWh)** | 5.376 | | | |
| **Portuguese demand contributors (%)** | 35% | | | |

**Table 4.** Comparison of actual values according to the IE and the second alternative counterfactual for: average wholesale electricity market prices, the contribution to demand made by the Iberian consumers concerned, the total cost to contributing consumers, average hourly Spanish congestion rents, the Spanish and Portuguese hourly demand and the % of demand concerned in each country; assuming France exported to Spain when actual French prices were lower than projected Spanish prices (15/06/22 - 23/09/22).

**5.3. Effects of the Iberian Exception**

Based on the scenario results summarized in Table 4, there are several effects of the Iberian Exception worth highlighting, which is done in the following subsections.

5.3.1 Effect on Spanish and Portuguese consumers in the Iberian market

The Iberian wholesale market prices in this paper's second counterfactual scenario are significantly higher than under the IE, but that price difference is less than the demand contributions borne by affected Iberian end consumers under the IE. Therefore, as shown in Table 4, the additional cost of the IE for affected Spanish consumers would have been 3%, or about €270 million[31], during the first 100 days. A sensitivity analysis that includes congestion rents in the counterfactual does not substantially affect the result[32]. This additional cost for affected consumers contrasts with the €1,400 million euros that the Ministry approach estimates as the savings for these consumers over the same period.

---

[31] From Table 4: 26,022 MWh x 59% x 7.3€/MWh x 2400 h = €269 million.
[32] A sensitivity analysis that includes the congestion rents as an immediate revenue source leads to a very small reduction in wholesale prices because congestion rents of €150 million are only 1% of the total cost of the wholesale market over the 100 days studied. The net effect of including these congestion rents in this counterfactual would be to increase the cost of the IE for affected consumer from 3% to 4%.

The case of Portugal is similar, but with a significantly lower percentage of consumers affected, this results in an additional IE cost for Portuguese consumers of around €33 million[33] during the first 100 days of the IE.

On the other hand, consumers with fixed-price contracts would not have seen their prices modified by the IE, so it had a neutral effect on them.

5.3.2 Effect on congestion rents

Congestion rents per kWh were equal to price differences between the wholesale prices on either side of the interconnector between France and Spain. These rents were shared 50/50 between the French and Spanish electricity systems. The application of the IE resulted in an increase in these rents relative to a counterfactual case without the IE, motivated by the IE's artificial reduction of wholesale prices in the Spanish market. The hourly average congestion rent for each country amounted to €251,000 under the IE compared to an estimated €64,000 had the IE not been introduced. That amounts to an increase in congestion rents of €450 million, from €150 million to €600 million, for each country over the IE´s first 100 days.

In this way, the French system saw its revenues from congestion rents increased by almost €450 million in the 100-day period considered. Normally, Spain would have used all its congestion rents to finance the costs of the system. However, presumably to win approval from the Commission and Portugal for the IE, Spain agreed to dedicate the entire €600 million from its congestion rents to finance the IE, thereby reducing the demand contribution for the affected consumers from Spain, Portugal, and Morocco[34].

Spain's share of the congestion rents was approximately €500 million and covered 13% of the costs of the IE over the 100 days. Were it not for those rents, the demand contribution of Spanish consumers would have risen by €17/MWh. Portugal´s share of the congestion rents over the same period saved Portuguese consumers subject to the mechanism almost €82 million[35]. Morocco obtained an €11 million reduction in its demand contribution [36].

5.3.3 Planet Earth

The IE mechanism assumes that CCGTs produce 0.55 MWh of electricity per MWh of natural gas. According to this paper's analysis, the IE increased Iberian electricity demand by 2.73 GWh, which is equivalent to an additional 6.5 TWh of gas that had to be imported.

---

[33] From Table 4: 5,376 MWh x 35% x 7.3€/MWh x 2400 h = €33 million.
[34] Note that the demand contributions were calculated after taking account of the congestion rents.
[35] From Table 4. Average hourly Portuguese demand (MWh)* Portuguese demand contribution (%) * Average reduction of DC due to congestion rents €/MWh * total hours = 5.376 MWh*35%*17 €/MWh*2400 = €82 million.
[36] Average hourly Moroccan demand (MWh) * Average reduction of DC due to congestion rents €/MWh * total hours = 278 MWh*17 €/MWh*2400 = €11 million.

This is probably an underestimate of incremental gas consumption because some of the incremental generation will have occurred in gas fired plants with lower efficiency.

Assuming $CO_2$ emissions of 0.38 tons per MWh and this paper's analysis of Iberian generation volumes, the IE caused an additional 577 tons of $CO_2$ to be emitted per hour in Iberia relative to our second counterfactual without the IE, which is equivalent to 1.4 million tons of $CO_2$ during the first hundred days of the IE mechanism. However, it is very difficult to know to what extent the increased emissions in Iberia increased global emissions since the authors do not know how France and Morocco would have met their demand if Spain had not exported to them.

The increased Iberian demand for gas may also have contributed to upward pressure on international prices of natural gas, but this too is very difficult to determine.

## 6. Suggestions for Further Research

The last section offered a summary of some of the quantified effects of the IE on electricity prices for affected Spanish and Portuguese consumers over the first 100 days of the IE relative to effects in two counterfactual scenarios developed in this paper. It also estimated the sharing of congestion rents among neighbouring countries as well as some of the other immediate effects, including higher $CO_2$ emissions and support for higher global gas prices.

However, these and other consequences of the IE deserve further research, as summarized below.

### 6.1 Immediate effects

6.1.1 Consumers

The impact of the IE on consumers requires further research. First, as with any counterfactual, the analysis presented in Section 5 reflects assumptions about how Spanish generators and consumers, as well as foreign electricity actors, would have behaved had the IE not been introduced, and refers only to the first 100 days. While these assumptions seem reasonable to the authors and more realistic than the Ministry's counterfactual, further analysis is certainly warranted, both within the same period and over longer periods. Sensitivity analysis is especially warranted to estimate the likely pricing and trading relationship between France and Spain. The authors presented reasons why they doubt that French wholesale prices would have been substantially different in the absence of the IE. However, it could be argued that the supply conditions in France, specifically the nuclear maintenance problems and drought, would have required France to import the same amount of electricity from Spain with or without the IE. To assess that claim requires the sort of market data for France that was used for Spain. It is also sensible to extend the analysis to cover a longer period, especially because the first 100 days of the IE included extremely high natural gas prices and, the operators and customers of the French power system hope, anomalous problems in the French generation sector.

Second, there is a need for further quantification of the demand contribution (DC) that financed the IE. It included two components, one related to compensation for generation in markets managed by OMIE and another for markets managed by REE. The analysis in this report includes only the compensation related to the OMIE markets. An initial assessment by the authors suggests that any savings or costs related to the REE markets do not materially change the conclusions. However, this requires further study.

Third, consumers with fixed price contracts were exempt from paying the DC at the outset of the IE. However, when they renewed what were often multi-year fixed price contracts, they were treated as if they were beneficiaries of the IE and thereby required to pay the DC. This often meant they were obliged to pay more than agreed in the multi-year fixed price contract. In that case, consumers that had hedged against price volatility were effectively being punished for doing so. Further study of the impact on these consumers is warranted. One hypothesis is that the government aimed to increase the volume of affected demand that was funding the IE to reduce the per unit cost, and it did so at a cost to the consumers who had signed multi-year fixed price contracts.

6.1.2. Generators

According to the conceptual analysis presented in Section 3, the IE had significant effects on generator margins. Further research is required to extend this analysis and to quantify the effects which are summarized below.

First, the privileged generators (combined cycle gas and coal generators) benefited from the IE because of its consequent increase in demand, largely due to increased French imports of power. Fossil-fired generators that would have been dispatched without the IE enjoyed increased revenues (market prices plus generation contributions) and margins under the IE. Most if not all the incremental fossil-fired generation that ran because of the increased demand resulting from the IE will also have obtained revenues and increased margins that otherwise they would not have enjoyed.

Second, the remaining (inframarginal) generators that were free to sell into the spot market may have seen their profits reduced due to lower wholesale market prices under the IE, without receiving the offsetting GC provided to privileged generators. However, much of this inframarginal generation (90 %) was not affected directly by the lower wholesale market prices because it was already subject to a price cap introduced before the IE came into effect.

Third, NERA has argued that the IE limited returns on regulated renewables generators and that this will later lead to a reimbursement and an increase in consumer costs. Although the authors of this paper do not think as a practical matter that the IE reduced renewables' returns below guaranteed levels, this issue may deserve further research. Further research may also be justified to assess the extent to which the IE reduced the returns for older renewables (mainly wind generation) that were no longer regulated.

### 6.2 Potential longer-term implications of the IE

It is difficult to be precise about the longer-term effects of the IE. However, some of the possible impacts that are detrimental to the energy transition can be identified and deserve further research.

First, following the standard economic framework introduced in Section 3, if the IE does reduce prices for final consumers or consumers think their prices are lower, the effect will be to reduce incentives to cut back on some of the energy consumption that consumers would have been willing to give up if they faced higher prices. This will discourage end-user investment in energy conservation and efficiency in use (e.g., heat pumps) and in technologies for shifting demand away from periods of peak prices. The result is to increase demand, perhaps especially at times of peak demand, or to reduce it by less than an efficient amount. This in turn contributes to additional fossil generation whose costs consumers (or taxpayers) will have to bear.

Second, any unexpected government intervention in the wholesale market, such as the IE, reinforces a well-deserved perception of regulatory risk, especially in Spain. On the supply side, the reduction of the wholesale price for inframarginal plants not covered by contracts or regulatory price caps almost certainly increases the perceived risk and cost of capital for investment in renewables and other inframarginal assets.

There is one possible longer-term effect of the IE that could potentially be beneficial in supporting the energy transition. To the extent that the IE signaled to consumers (probably erroneously) a sustainable reduction in the end-user price of electricity, the IE intervention to reduce prices might have encouraged electrification.

## 7. Concluding Comments on the Research

This paper summarizes the authors' initial assessment of the impact of the Iberian Exception on non-exempt Spanish consumers in its first 100 days. First, it questions the Spanish Government´s claims that the IE reduced affected consumer prices by up to 20%. The analysis in this paper suggests that the energy component of affected consumer prices fell by less than that and that, based on this paper's scenario reflecting the effects of the IE on French power imports, affected consumers might have paid less if the IE had not been introduced. A key reason for the difference in views is that the authors' methodology takes changes in both demand and supply into account.

Second, the conceptual analysis in Section 3 suggests that the IE increased fossil generator margins and diminished margins for decarbonized generation. This is an important issue since it is presumably not the objective of the IE. Further research is required to assess that hypothesis and to quantify its effects.

Finally, although further analysis of the IE exception is certainly warranted, it is hard to escape a general conclusion, namely that intervention in electricity markets introduces many distortions, some predictable and others perhaps not, that have a damaging effect on the energy transition. Perhaps the greatest concern is related to distorted prices signals: suggesting to consumers that they need not contract to hedge price volatility and invest to manage their demand; and suggesting to upstream investors that they should adopt a higher cost of capital to reflect regulatory risk in Spain. These signals are especially problematic now, as Spain, Portugal, and the rest of the EU commit to a transition that requires massive investments by consumers and industry throughout the energy system.